\documentstyle[eqsecnum,aps,prd,epsfig]{revtex}
\date{\today}
\draft
\def\be{\begin{equation}}
\def\ee{\end{equation}}
\def\beu{\begin{displaymath}}
\def\eeu{\end{displaymath}}
\def\bea{\begin{eqnarray}}
\def\eea{\end{eqnarray}}
\def\beau{\begin{eqnarray*}}
\def\eeau{\end{eqnarray*}}

\def\n2{\langle N^2 \rangle}
\def\sn2{\sqrt{\langle N^2 \rangle}}
\def\laq{\raise 0.4ex\hbox{$<$}\kern -0.8em\lower 0.62
ex\hbox{$\sim$}}
\def\gaq{\raise 0.4ex\hbox{$>$}\kern -0.7em\lower 0.62
ex\hbox{$\sim$}}

\def\L{{\cal L}}
\def\sframe{\hbox{$\underline{S\ }\!\!|$\ }}
\def\eframe{\hbox{$\underline{E\ }\!\!|$\ }}

\def\half{\hbox{\small $\frac{1}{2}$}}
\def\NPB{{\em Nucl. Phys.} B}
\def\PLB{{\em Phys. Lett.}  B}
\def\PRL{{\em Phys. Rev. Lett.}}
\def\PRD{{\em Phys. Rev.} D}
\def\MPL{{\em Mod. Phys. Lett.}  A}
\newcommand{\figref}[2]{Fig. \ref{#1}#2}
\newcommand{\bfig}{\begin{figure}}
\newcommand{\efig}{\end{figure}}
\def\figwidth{17cm}
\def\figheight{15cm}

\newcommand{\bigeps}[1]{\begin{center}\epsfig{file=#1,width=\figwidth,
                        height=\figheight,bbllx=60,bblly=40,
                        bburx=560,bbury=470}\end{center}}
\newcommand{\littleeps}[1]{\begin{center}\epsfig{file=#1,width=7cm, 
                        height=7cm,bbllx=12,bblly=-20,
                        bburx=576,bbury=480}\end{center}}

\begin{document}
% 
%%%%%%%%%%%%%%%%%%%%%%%%%%%%%%%%%%%%%%%%%%%%%%%%%%%%%%%%
%
\title{A Model of Graceful Exit in String Cosmology}

\author{Ram Brustein, Richard Madden}
\address{Department of Physics,
Ben-Gurion University,
Beer-Sheva 84105, Israel\\
email: ramyb@bgumail.bgu.ac.il, madden@bgumail.bgu.ac.il }

\maketitle
\begin{abstract}
We construct, for the first time, 
a model of graceful exit transition from a dilaton-driven inflationary phase
to a decelerated Friedman$-$Robertson$-$Walker era. 
Exploiting a demonstration 
that classical corrections can stabilize a high curvature string 
phase while the evolution is still in the weakly coupled regime, 
we show that if additional terms of the type that may result from
quantum corrections to the string effective action exist, 
and induce violation 
of the null energy condition, then  
evolution towards a decelerated Friedman$-$Robertson$-$Walker
phase is possible.  
We also observe that stabilizing the dilaton at a fixed value,
either by capture in a potential minimum or by radiation production, 
may require that these quantum corrections
are turned off, perhaps by non-perturbative effects or higher order 
contributions which overturn the null energy condition violation.
\end{abstract}

\vskip -0.1in
\centerline{Preprint Number: BGU-PH-97/11}
\section{INTRODUCTION}

An inflationary scenario \cite{gv1,gv2}, inspired 
by duality symmetries of string
cosmology equations \cite{gv1,dual,tv}, is based on the fact that
cosmological solutions to string dilaton-gravity 
come in duality-related pairs,
the plus branch $(+)$, and the minus branch  $(-)$ \cite{bv}. The $(+)$ branch
has kinetic inflationary solutions in which the Hubble parameter increases with
time. The minus branch  $(-)$  can be connected smoothly to a standard
Friedman$-$Robertson$-$Walker (FRW) decelerated  expansion of the Universe with
constant dilaton. The scenario (the so called ``pre-big-bang" scenario) is that
evolution of the Universe starts from a state of very small curvature and
coupling and then undergoes a long phase of dilaton-driven kinetic inflation
described by the $(+)$ branch and at some later time joins smoothly standard 
radiation dominated cosmological evolution, thus giving rise to a singularity
free inflationary cosmology. Recently, the required initial
conditions of the Universe in this scenario were discussed \cite{inhom}. Our
focus in this paper is mainly on the final fate of the Universe at late
times and we simply assume that the correct initial conditions were chosen
such that a long dilaton-driven inflationary phase was indeed part of the
evolution.

The graceful exit transition from the initial long phase  of
dilaton-driven kinetic inflation to the subsequent standard radiation dominated
evolution has been a subject of many investigations 
\cite{BM,classexit,quantumexit,gmv}. In \cite{bv} it was argued,  
and later proved
\cite{kmo}, that such a transition cannot occur while  curvature was below the
string scale and the string coupling was still weak, leading to the conclusion
that an intermediate ``string phase" of high curvature 
(previously  suggested as
a possibility \cite {gv1,gv2})  or strong coupling is actually required
\cite{bggv}.  

In \cite{BM} we proposed to use an effective description in terms of sources
that represent arbitrary corrections to the lowest order equations. We were 
able to relate necessary conditions for graceful exit to energy conditions
appearing in singularity theorems of Einstein's general relativity \cite{he}. 
In particular, we showed that a successful exit requires violations of the
null energy condition (NEC). Since most classical sources obey NEC this
conclusion hints that quantum effects, 
known to violate NEC in some cases, may be
the correct sources to look at. To briefly recap the relevant results of
\cite{BM} we need to recall that there are different conformal frames in which
to describe the equations. 
These are related by local field redefinitions which, 
supposedly, do not affect physical observables \cite{gv3}. The two frames that
we use are the string frame (\sframe) and  the ``lowest order Einstein frame"
(\eframe) in which the lowest order classical action kinetic terms of gravity
and the dilaton are diagonal. For brevity we call \eframe the Einstein frame. 
The general analysis of \cite{BM} resulted in a set of necessary conditions
on the evolution in terms of the Hubble parameters 
$H_S$ in the string frame and
$H_E$ in the Einstein frame and the dilaton $\phi$.
We include them for completeness 
\begin{itemize}
\item
Initial conditions of a (+) branch and $H_S,\dot\phi>0$  require
$H_E<0$.
\item
A branch change from (+) to $(-)$ has to occur while $H_E<0$.
\item
A successful escape and exit completion requires NEC violation
accompanied by a bounce in \eframe
after the branch change has occurred, ending up with $H_E>0$.
\item
Further evolution is required to bring about a radiation
dominated era in which the dilaton effectively
decouples from the ``matter" sources.
\end{itemize}

The question we set to answer in this paper is the following. Suppose that
effective sources of the type that are expected to appear as 
corrections to the lowest order effective action of strings 
do provide NEC violation. Then
would a complete exit transition actually occur?
We answer this question in the affirmative, but some surprising obstacles 
are found on the way and we also suggest ways to overcome these
obstacles. 

Because in our scenario the Universe evolves towards higher curvatures and
stronger coupling,  there will be some time when the lowest order effective
action can no longer reliably describe the dynamics and it must be corrected.
Corrections to the lowest order effective action come from two sources. The
first are classical corrections, due to the finite size of strings, arising
when the fields are varying over the string length scale
$\lambda_s=\sqrt{\alpha'}$. These terms are important in the regime of large
curvature.  The second are quantum loop corrections. The loop expansion is
parameterized by powers of the string coupling parameter $e^\phi=g_{string}^2$,
which in the models that we consider is, of course,  time dependent.  So
quantum corrections will become important when the dilaton becomes large, the
regime we refer to as strong coupling.   

In this paper we make use of both types of corrections. The role of $\alpha'$
corrections \cite{gmv} is to create an attractive fixed point which stabilizes
the evolution in a high curvature regime with linearly growing dilaton. The
basin of attraction of this fixed point is large enough to allow it to be
reached from generic initial conditions. Further, the location of this fixed
point is such it forces the evolution to undergo a branch  change, and all of
this may occur for small values of the dilaton (weak coupling) so the quantum
corrections can be ignored.  With the linearly growing dilaton, the quantum
corrections will  eventually become important. It is these we will attempt to
use to let the Universe escape the fixed point and complete the  transition to
a decelerated FRW evolution. We allow ourselves the freedom to choose the
coefficients of the quantum correction terms arbitrarily, in particular, their
sign is chosen so as to induce NEC violation. Our reasoning for allowing this
freedom stems in part from a lack of any real string calculations and in part
by our desire to verify by constructing explicit examples the general arguments
of \cite{BM}. 

We use the general framework set up in, and rely for analytical
calculations mostly on \cite{BM}. We perform numerical integration of specific
equations with limited new analytical considerations. 
We show, by constructing explicit examples, that a complete exit transition is
not forbidden, which, of course, does not yet prove that in string theory it
actually occurs. For that, an explicit string computation of quantum correction
terms is necessary. This may also require taking into account the backreaction
of the particles produced during the dilaton-driven phase.

A by-product of our analysis is a detailed description of the
high-curvature (``string'') phase in between the dilaton-driven 
inflationary phase and the decelerated FRW phase. 
We obtain a more complicated
phase than the one postulated in \cite{bggv} with non-constant $\dot\phi$ and
$H_S$. We do not take the details of the evolution in the string phase too
seriously, because the terms that determine those details have arbitrary
coefficients. However, our examples could be taken as an indication of what
the real string phase may eventually look like.

The paper is organized as follows. In section II we present 
and discuss solutions
which exhibit branch change and exit completion with $\alpha'$ and
quantum corrections. In section III we explain the remaining problem of
correction dominated evolution, show how to model a shut-off of the corrections
and verify that once corrections are shut off, 
the usual decelerated evolution such
as radiation domination or capturing of the dilaton in a minimum of a
potential may follow. We summarize our results and explain which
further calculations are necessary to reach a conclusive statement about exit
transition in string models. In a technical appendix we outline details of
derivation of equations of motion, numerical integration etc.

\section {A Model for Exit Completion}
 
String theory effective action in four dimensions 
takes the following general form, 
\begin{equation} 
S_{eff}^{{\sframe}}= 
\int d^4 x \left\{ \sqrt{-g}\left[ \frac{e^{-\phi}}{16 \pi \alpha'} 
\left(R+\partial_\mu\phi \partial^\mu\phi\right)\right]+ 
\half \sqrt{-g} \L_c\right\}, 
\label{effacts} 
\end{equation} 
where $g_{\mu\nu}$ is the 4-d metric and $\phi$ is the dilaton,  
the effective action is written here in the string frame (\sframe).  
The Lagrangian $\L_c$ may contain arbitrary corrections to the  
lowest order 4-d action coming from a variety of sources as 
specified below.  
 
We are interested in solutions to  the equations of motion derived from  
the action (\ref{effacts}) of the FRW type with vanishing spatial curvature 
(non-vanishing spatial curvature may be included as an effective source in 
the equations) 
$ds^2= -dt_S^2+a_S^2(t) dx_i dx^i$ and $\phi=\phi(t)$.  
We will assemble the equations of motion by 
deriving the correction energy-momentum  
tensor $T_{\mu\nu}=\frac{1}{\sqrt{-g}} 
\frac{\delta \sqrt{-g} \L_c}{\delta g^{\mu\nu}}$,  
which will have the form $T^\mu_{\ \nu}=diag(\rho,-p,-p,-p)$. In addition 
we have another form of source term arising from the variation by $\phi$  
equation, $\Delta_\phi\L_c=\half \frac{1}{\sqrt{-g}}   
\frac{\delta \sqrt{-g} \L_c}{\delta\phi}$. 
 
In terms of these sources the equations of motion are 
\begin{eqnarray} 
3H_S^2+\half\dot\phi^2-3 H_S\dot\phi&=&  
\half e^{\phi} \rho \label{n00eq} \\ 
-2 \dot H_S -3 H_S^2 + 2 H_S\dot\phi - \half \dot\phi^2 + \ddot\phi&=&  
\half e^{\phi} p \label{n11eq} \\  
3 \dot H_S + 6 H_S^2 - 3 H_S\dot\phi + \half \dot\phi^2 - \ddot\phi&=&  
\half e^{\phi} \Delta_\phi\L_c \label{phieq} \\  
\dot\rho+3 H_S(\rho+p)&=&  
-\Delta_\phi\L_c \dot\phi, \label{nconseq} 
\end{eqnarray} 
$H_S=\dot a_S /a_S$, and we have fixed  our 
units such that $16\pi\alpha'=1$.  
 
We will make the split 
\begin{equation} 
\L_c=\L_{\alpha'}+\L_{q}+\L_{m} \\ 
\label{actsplit} 
\end{equation}  
and we will divide up the source terms analogously. For example, 
$\rho=\rho_{\alpha'}+\rho_{q}+\rho_{m}$. We take, for concreteness, the Lagrangian 
$\L_{\alpha'}$ to be of the specific form of the $\alpha'$ classical 
corrections proposed in \cite{gmv}, 
\begin{equation} 
\half \L_{\alpha'}=e^{-\phi}
                   (\frac{R_{GB}^2}{4}- \frac{(\nabla \phi)^4}{4}) .
\label{gmvlag}
\end{equation} 
The contribution of this term to the effective sources is  
\begin{eqnarray} 
\rho_{\alpha'}&=&e^{-\phi} (6H_S^3 \dot \phi-\frac{3 \dot \phi^4}{4})  \\ 
p_{\alpha'}&=&e^{-\phi} (-4 H_S^3 \dot \phi-4H_S \dot H_S \dot \phi 
+2 H_S^2 \dot \phi^2-\frac{\dot \phi^4}{4}-2 H_S^2 \ddot \phi) \\ 
\Delta_\phi \L_{\alpha'}&=&e^{-\phi} (-6 H_S^4 -6 H_S^2 \dot H_S+3 H_S  
\dot \phi^3-\frac{3 \dot \phi^4}{4}+3 \dot \phi^2 \ddot \phi).  
\end{eqnarray} 
Although we have fixed our units differently from \cite{gmv} (where
$k \alpha'=1$, k depending on the species of string theory), we have
chosen the coefficient of this term so as to lead to solutions which are
numerically identical to theirs for purposes of comparison.

The Lagrangian $\L_{q}$ will designate quantum loop corrections parameterized 
by powers of $e^{\phi}$. We will examine a variety of candidates 
for $\L_{q}$. The Lagrangian $\L_{m}$ will represent such things as radiation  
or a dilaton potential  when we discuss the final transition to  
radiation domination and a fixed dilaton. 
 
Setting $\L_{q}=\L_{m}=0$, for the moment, we integrate the equations of 
motion starting from initial conditions near the $(+)$ branch vacuum. 
To do this we 
solve (\ref{n11eq}) and (\ref{phieq}) for the highest derivatives of $H_S$ and 
$\phi$ and use (\ref{n00eq}) as a constraint on the initial conditions. 
The conservation equation (\ref{nconseq}) is an identity owing to the 
usual redundancy in the equations, only when we add radiation to the 
system will we use this equation in the evolution. 
We recover the results of \cite{gmv} in four 
dimensions where they find attraction into a fixed point at $H_S=0.616..., 
\dot \phi=1.404...$. 

As shown in \figref{f:gmv}{a} the solution begins near 
the $(+)$ branch vacuum and makes a branch change  
but does not ``complete'' the exit by proceeding to the 
$\rho>0$ region. The quantity characterizing branch sign 
is plotted in \figref{f:gmv}{e}. We examined more general 
forms of this Lagrangian involving different coefficients.
Varying the coefficients moves the fixed point. For some ratios of 
coefficients the solutions encounter singularities before reaching
the fixed point (we discuss the form of these singularities in section
3) but other ratios yield solutions of the same 
generic type as we have presented here.

In spite of considerable progress we still 
need to look more closely at the sources and Einstein frame  
evolution to answer the question 
of how close we have come to the true goal of decelerated FRW 
evolution. 
 
The evolution of $\rho$ and $p$ is easily found by substituting the 
solution back into the equations of motion and solving for the sources. 
We see that this phase does violate NEC in the string frame. However, 
recalling our stronger claim that there should also be NEC violation 
in the ``lowest order Einstein frame'' (denoted with a 
subscript $E$ and as explained in the introduction, many times for 
brevity called just the Einstein frame), 
we can compute the value of the Einstein frame sources by finding the Hubble 
parameter $H_E$ and its  derivative $\dot H_E$,
\begin{eqnarray} 
H_E &=& e^{\phi/2} (H_S-\half \dot \phi) \\ 
\dot H_E &=& e^{\phi} (\dot H_S - \half \ddot \phi + \half \dot \phi H_S 
                       -\frac{\dot \phi^2}{4}) 
\end{eqnarray} 
All quantities on the right hand side of these are given strictly in 
terms of string frame quantities and time. 
The Einstein equations are, 
\begin{eqnarray} 
3 H_E^2&=&\half \rho_E \\ 
3 H_E^2+2 \dot H_E &=& -\half p_E 
\end{eqnarray} 
Notice that we have absorbed the dilaton kinetic energy into the definition 
of $\rho_E$ and $p_E$. 
 
\bfig
\bigeps{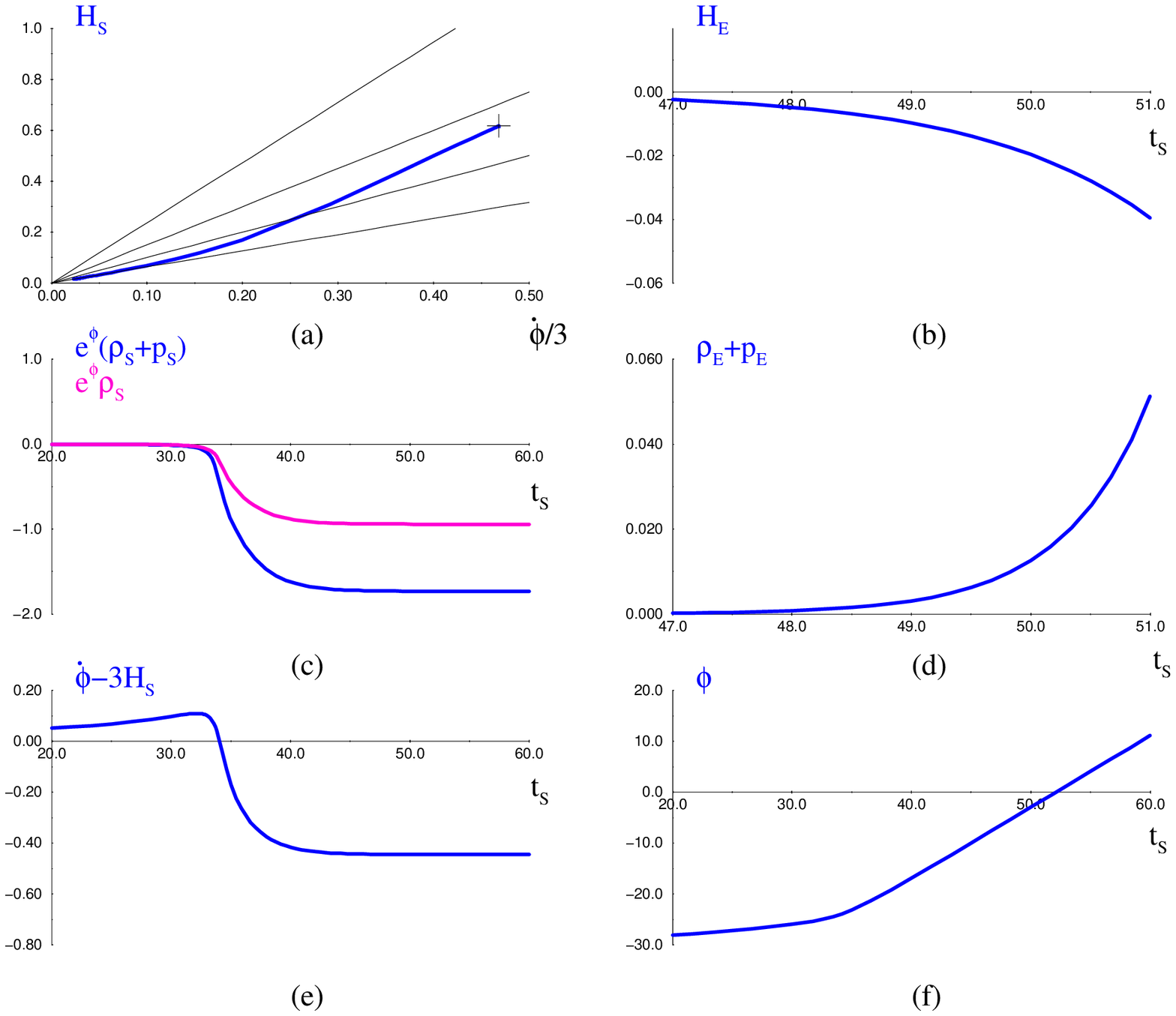}
\caption{ 
\it \small 
A solution to the classical equations with only $\alpha'$ corrections, 
$\L_q= \L_m=0$. (a) Evolution 
in the $(\dot \phi/3,H_S)$ plane. The four lines 
plotted are, in order of increasing slope, the $(+)$ branch vacuum  
($\rho=0, \dot \phi=(3+3^{1/2}) H_S$), 
branch change ($\dot \phi=3 H_S$), 
Einstein frame ``bounce'' ($\dot \phi=2 H_S$) and $(-)$ branch   
vacuum ($\rho=0, \dot \phi=(3-3^{1/2}) H_S$).  
The cross is at the location of the fixed point. The remaining figures 
are various quantities plotted as a function of string time. 
(b) The Einstein frame 
Hubble parameter $H_E$. (c) The string 
frame source terms $e^\phi (\rho_S+p_S)$ and $e^\phi \rho_S$. 
(d) The Einstein frame source term $\rho_E+p_E$. (e) $\dot \phi-3 H_S$, 
a quantity indicating branch sign. (In section III this figure will 
be replaced by $e^\phi \rho_S$ from (c)), 
(f) $\phi$ evolution. Initial conditions 
at $t=0$, $\phi=-30, H_S=0.0148529, \dot \phi=0.0699602$. The following 
figures share these initial conditions, since quantum corrections are 
very small at this time. The $\phi$ evolution may be started at 
arbitrarily small values by evolving these initial conditions further 
backward towards the $(+)$ branch vacuum. }  
\label{f:gmv}
\efig 

The resulting sources and $H_E$ are plotted in \figref{f:gmv}{b,c,d}. 
From the figures we see that there is no violation of NEC in the Einstein 
frame, corresponding to the fact that there is no ``bounce''. This 
solution represents a singular collapse in the Einstein frame because  
of the linearly increasing dilaton plotted in \figref{f:gmv}{f}. In terms 
of sources this suggests that there is insufficient NEC violation 
and that the addition 
of conventional sources to $\L_m$, like radiation, which do not violate  
NEC cannot help the completion of the exit transition.
 
\bfig 
\bigeps{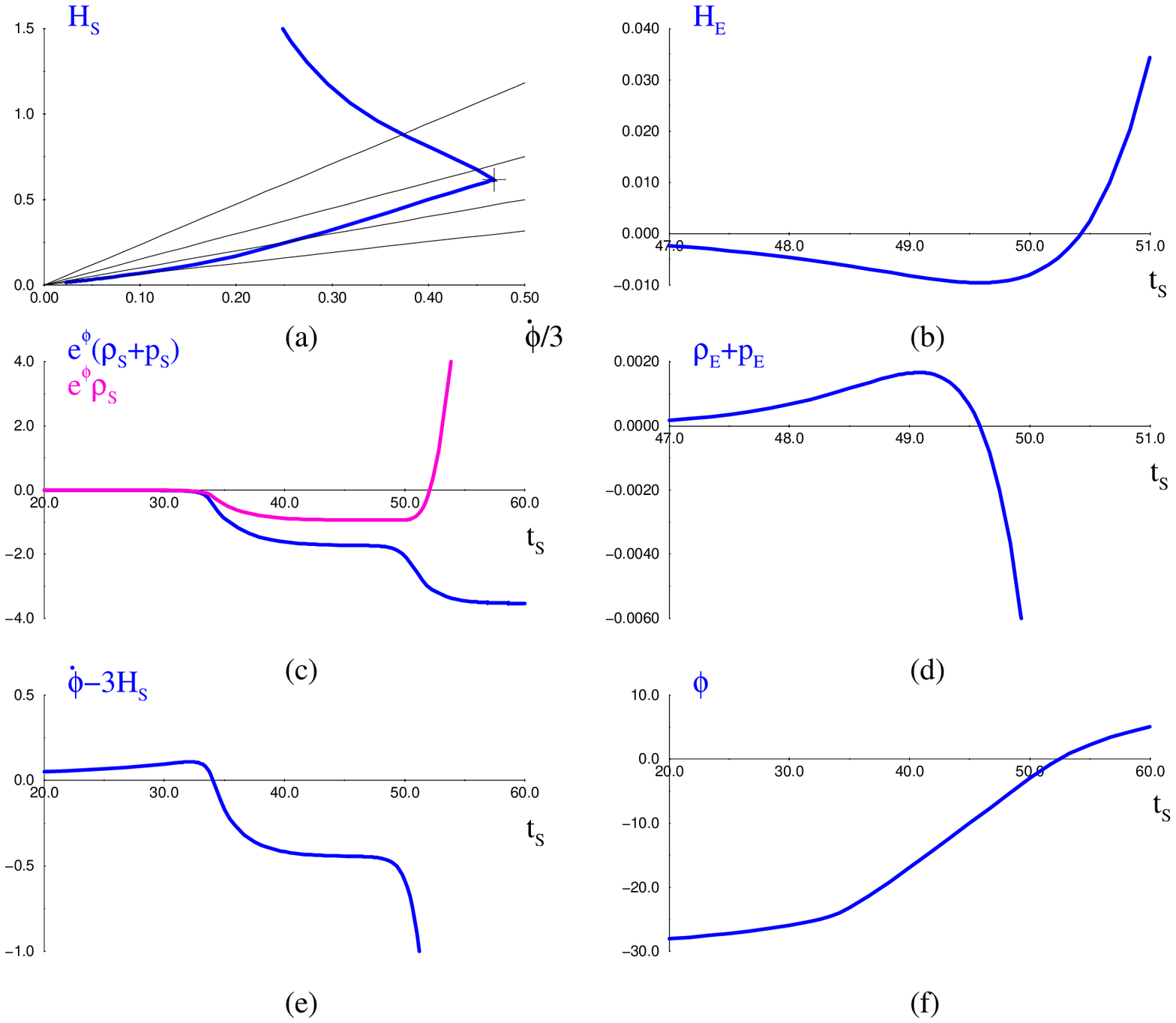}
\caption 
{\small\it 
$\half \L_q^{\phi}=-(\nabla \phi)^4$, $\L_{m}=0$. See \figref{f:gmv}{} 
caption for details and initial conditions. } 
\label{f:gp4} 
\efig 
 
Perhaps generically any correction violating NEC strongly 
enough will complete the transition. In particular, we consider terms 
modeling the form of quantum loop corrections parameterized by powers 
of $e^{\phi}$. For example, at one loop, 
\begin{eqnarray} 
\half \L_q^{\phi}&=&- (\nabla \phi)^4 \label{gp4}\\ 
\rho_q^{\phi}&=&-3 \dot \phi^4 \\ 
p_q^{\phi}&=&- \dot \phi^4 \\ 
\Delta_{\phi} \L_q^{\phi}&=&12 (H_S \dot \phi^3 + \dot \phi^2 \ddot \phi) 
\end{eqnarray} 
As long as we include only one order of loop correction, 
the overall coefficient of our corrections
can be absorbed by a shift of $\phi$, and it therefore determines the
value of $\phi$ at which the quantum corrections begin to be important,
but doesn't lead to qualitatively different behavior.
So at this stage we choose coefficients of order unity and after
these preliminary investigations are complete, we
will choose more realistic values.

Clearly, because of the factor of $e^\phi$ on the right hand side of 
equations (\ref{n00eq}), (\ref{n11eq})
this $\L_q^{\phi}$ can give us NEC violation of increasing strength as $\phi$ 
increases. We show the results of the numerical integration 
by presenting the same suite of figures 
as in the previous case in \figref{f:gp4}{}. 
In this case we have chosen to graph the sources for a range of time 
emphasizing the Einstein frame bounce. 
As hoped, the solution now proceeds into the $\rho>0$ region triggered 
by increased NEC violation in the string frame. 
We also see the accompanying Einstein frame NEC violation and bounce.  
Notice that the bounce occurs quickly, during a few units of time, 
compared with the very long duration of the dilaton-driven phase 
near the $(+)$ vacuum.

However, inspecting the solution at late times shows the dominant 
terms in the equations of motion are: 
\begin{eqnarray} 
-6 H_S^2-3 e^{\phi} \dot \phi^4+6 H_S^3 \dot \phi &=& 0 \\ 
-6 H_S^2+e^{\phi} \dot \phi^4+4 H_S^3 \dot \phi &=& 0 \\ 
-6 H_S^4+12 e^{\phi} H_S \dot \phi^3 &=& 0 
\end{eqnarray} 
This system has the solution,  
\begin{eqnarray} 
H_S&=&{2 t \over 3} \\  
\phi&=&\log ( { t^6 \over 39366} ) 
\end{eqnarray} 
approximating the nature of the true solution at late times.  
 
This form confirms that the solution has unbounded growth in the 
curvature and dilaton. Nonetheless, we have succeeded in our aim  
of completing the exit to the $\rho>0$ region of phase space. 
It may appear that we are now facing a new ``graceful exit'' problem 
since the equations of motion at late times are dominated by corrections,  
spoiling the expected stability of a $(-)$ branch. 
We suggest that the source of this instability 
is the continued NEC violation and will search for a cure in the 
next section.  
 
To further explore the sufficiency of NEC violation for exit  
we examine another generic form of one loop  
quantum correction, 
\begin{eqnarray} 
\L_q^{R^2}&=& R^2 \\ 
\rho_q^{R^2}&=& -108 H_S^2 \dot H_S+18 \dot H_S^2-36 H_S \ddot H_S \\ 
p_q^{R^2}&=& 108 H_S^2 \dot H_S+54 \dot H_S^2 + 
             72 H_S \ddot H_S+12 H_S^{(3)} \\ 
\Delta_{\phi} \L_q^{R^2}&=& 0 
\end{eqnarray} 
We immediately notice that for evolution strictly in the fixed point,  
where $H_S=constant$, this correction will be zero indicating it 
will not effectively contribute to NEC violation. 
Furthermore, this introduces higher  
derivatives into the equations of motion, which in turn introduces  
dangerous pathologies into the solutions \cite{wald}. 
 
These pathologies come from the extra degrees of freedom in the system 
coming from the extra initial conditions that need to be imposed. These 
extra degrees of freedom are associated with unstable modes of the  
solutions which we regard as physically spurious, since they are solutions 
in which the ``correction'' becomes much larger than the terms to which 
it is supposed to be a small correction. Reference \cite{wald} suggests two  
remedies to this problem. The first is ``reduction of order'', in which 
we differentiate the large terms in the equations of motion (in this case 
the tree level terms) and use them to rewrite the higher derivatives in 
terms of lower derivatives, justified by the assumption the corrections 
will remain small. The second is simply to carefully chose initial  
conditions to avoid the unstable modes. 

Since our purpose is to make a qualitative survey of the effect of 
corrections and the reduction of order prescription is computationally
prohibitive, we take the second approach and choose initial conditions 
so that the evolution is not dominated by the corrections for a 
reasonable span of time. Exploring these solutions confirms 
that $\L_q^{R^2}$ corrections themselves do not help the exit process.
The other one loop curvature squared terms, $R^{\mu \nu} R_{\mu \nu}$
and $R^{\mu \nu \sigma \rho} R_{\mu \nu \sigma \rho}$ yield sources
which are constant multiples of those of $R^2$, and as is well known,
the Gauss-Bonnet combination, which does not contribute higher derivatives,
vanishes in the equations of motion.

But putting curvature squared terms together with corrections 
that do not vanish in the fixed point, for example  
$\L_q=\L_q^\phi-\frac{\L_q^{R^2}}{3}$, does yield qualitatively
different solutions which do exit, 
in this case a solution approaching string frame deSitter (with 
constant $H_S$), though again with still growing dilaton as  
illustrated in \figref{f:r2gp4}{}. We set our initial conditions 
near a later phase of the $(+)$ branch vacuum to avoid instabilities. 
 
\bfig 
\bigeps{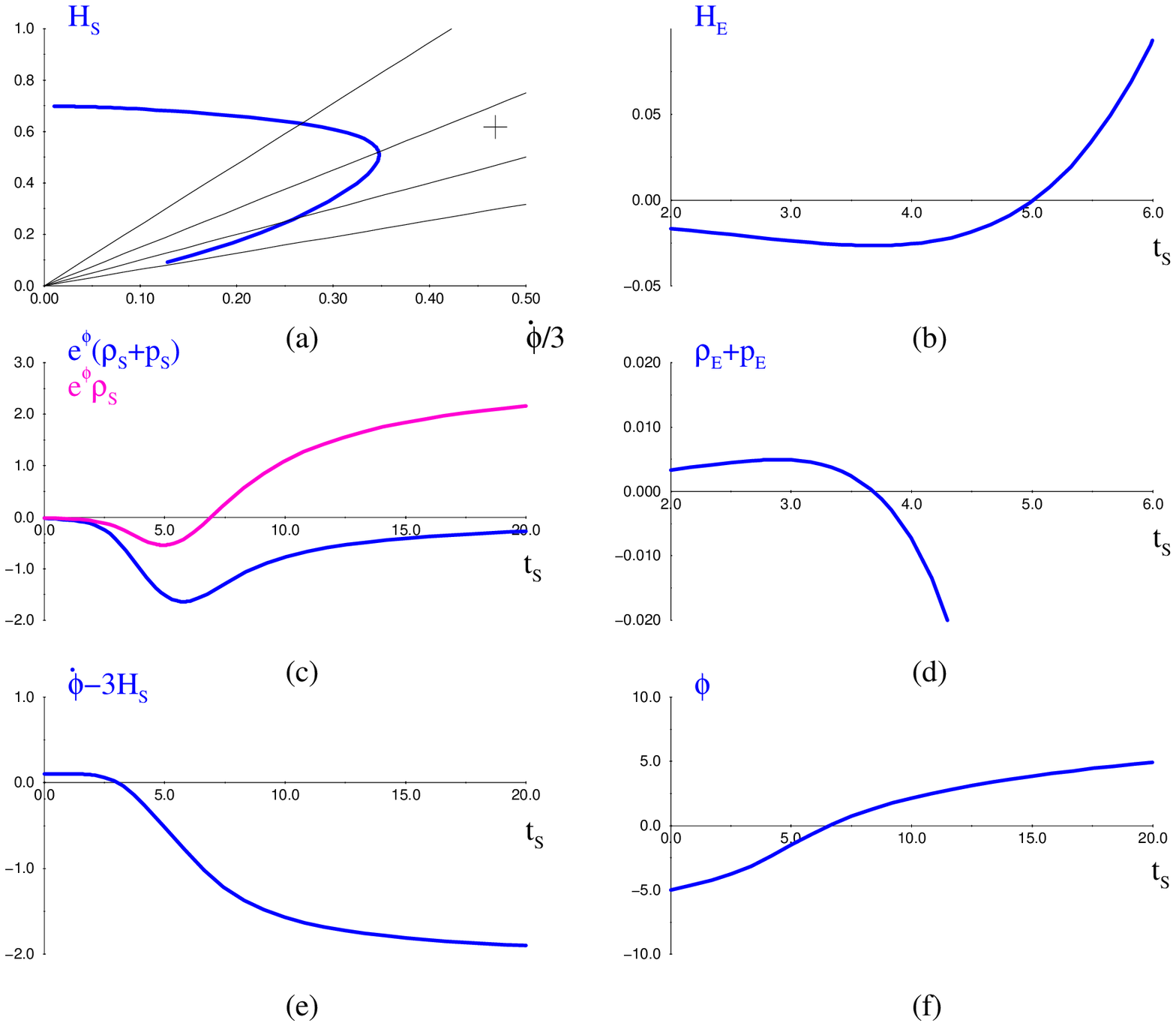}
\caption{ 
\small \it 
$\L_q=\L_q^\phi-\frac{1}{3}\L_q^{R^2}$, $\L_{m}=0$. Initial conditions 
at $t=0$, $\phi=-5$, $H_S=0.0926031$, $\dot \phi=0.383925$, 
$\dot H_S=0.018527$, $\ddot H_S=0.0260335$. See \figref{f:gmv}{} caption 
for explanation. }
\label{f:r2gp4} 
\efig 
 
Finally, we present a two loop Gauss-Bonnet correction which we choose 
because it represents the influence of curvature squared terms but  
does not contribute higher derivative terms. 
\begin{eqnarray} 
\half \L_q^{R^2_{GB}}&=& e^{\phi} R^2_{GB} \\ 
\rho_q^{R^2_{GB}}&=& e^{\phi} (-24 H_S^3 \dot \phi) \\ 
p_q^{R^2_{GB}}&=& e^{\phi} (16 H_S^3 \dot \phi + 16 H_S \dot H_S \dot \phi 
                +8 H_S^2 \dot \phi^2 + 8 H_S^2 \ddot \phi) \\ 
\Delta_{\phi} \L_q^{R^2_{GB}}&=& e^{\phi} (24 H_S^4+24 H_S^2 \dot H_S) 
\end{eqnarray} 
Again we find that introducing this correction with an appropriate sign 
can complete the transition to $\rho>0$ as in \figref{f:l2gb}{},  
but with increasing domination by the 
correction terms leading to a singularity soon after the transition.  
We have also tried other combinations of sources with different coefficients 
and found that many of them yield solutions that are similar to the ones 
we presented, 
leading us to believe that our result are quite general and do not depend in 
a strong way on particular initial conditions or coefficients.
 
\bfig 
\bigeps{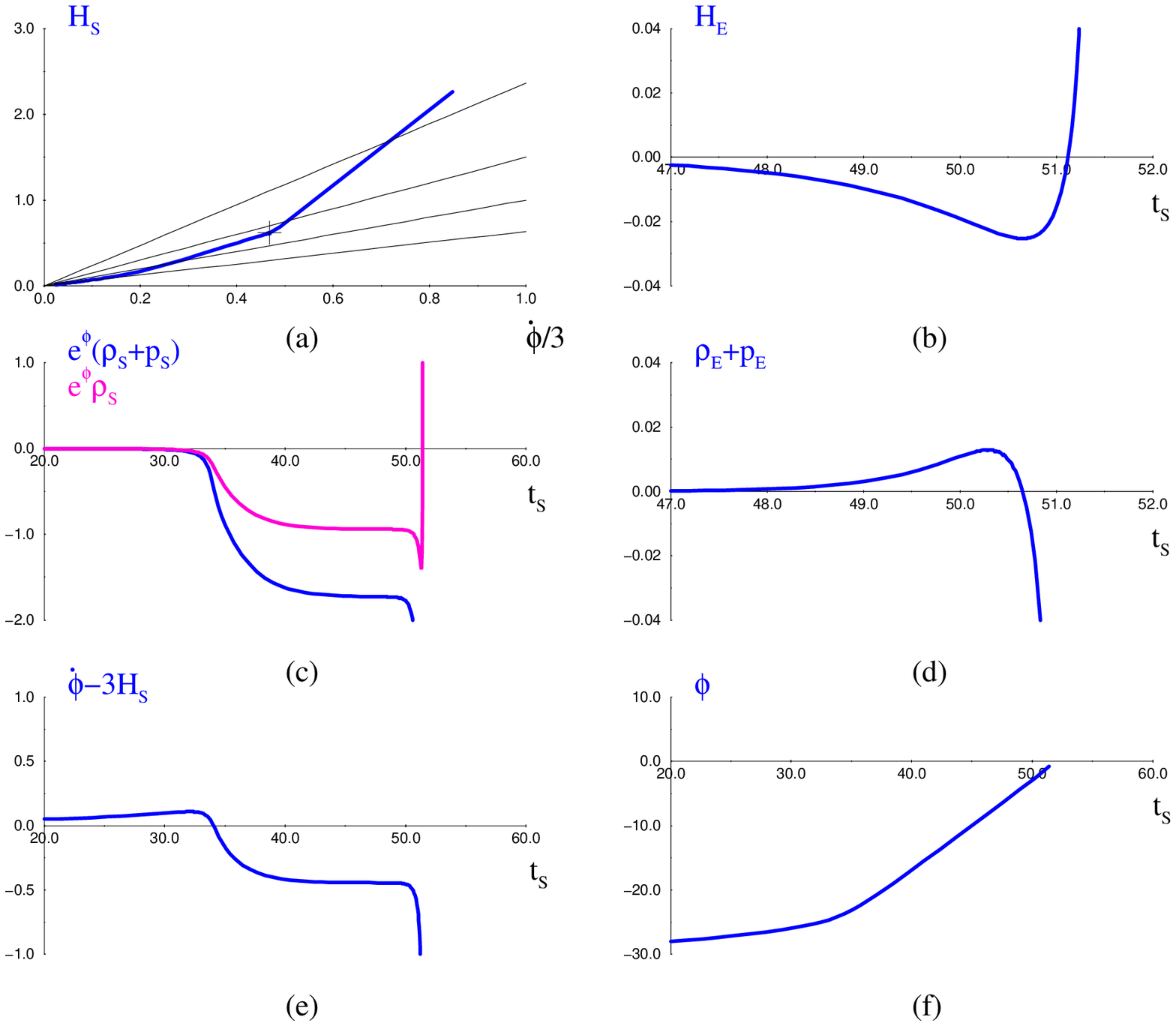}
\caption 
{\small\it 
$\half \L_q^{R^2_{GB}}=e^{\phi} R^2_{GB}$, $\L_{m}=0$. See \figref{f:gmv}{} 
caption for details and initial conditions.} 
\label{f:l2gb} 
\efig 
 
In summary, we have seen that generic forms of 
quantum corrections can complete 
the exit from the fixed point of \cite{gmv} to the region of $\rho>0$, 
showing the NEC violation is not only necessary, but is in some sense 
sufficient. The resulting solutions are quite varied,  
but we have also noted they have 
unbounded growth of the dilaton at late times,   
continue to be dominated by corrections and continue to violate NEC. 
In spite of being $(-)$ branch solutions they are still unstable. 
However, inducing the exit seems to be a generic property of 
NEC violation and not of the specific form of the corrections. 
In the next section we will attack this new exit problem, the exit from 
the epoch of correction domination. 
 
\section { A Model for Transition to Decelerated and Stable Evolution } 
 
We have seen that by using plausible forms for quantum  
corrections we can induce NEC violation and 
push the evolutions into a region we would like to call a completed exit. 
However, we have also seen that these evolutions are dominated by the 
corrections and have singular behavior unlike the desired $(-)$ branch 
solutions. We associate this behavior with two overlapping sources. 
First is the continued NEC violation itself, which tends to feed  
accelerated evolution in the Einstein frame. Second, and more directly, it 
is the continuing growth of $\phi$ that supports the strength of the  
quantum correction terms through the powers of $e^\phi$ occurring in the 
equations of motion. 
 
Since these solutions do not spontaneously suppress the corrections we  
might hope that simply controlling the growth of the dilaton would tame 
the solutions. This can be done by modeling radiation production, which  
can slow the change in the dilaton through the $3 H_S \dot \phi$ ``friction''  
term in (\ref{phieq}), or more directly by capturing the 
dilaton in a potential minimum. 
 
\bfig
\littleeps{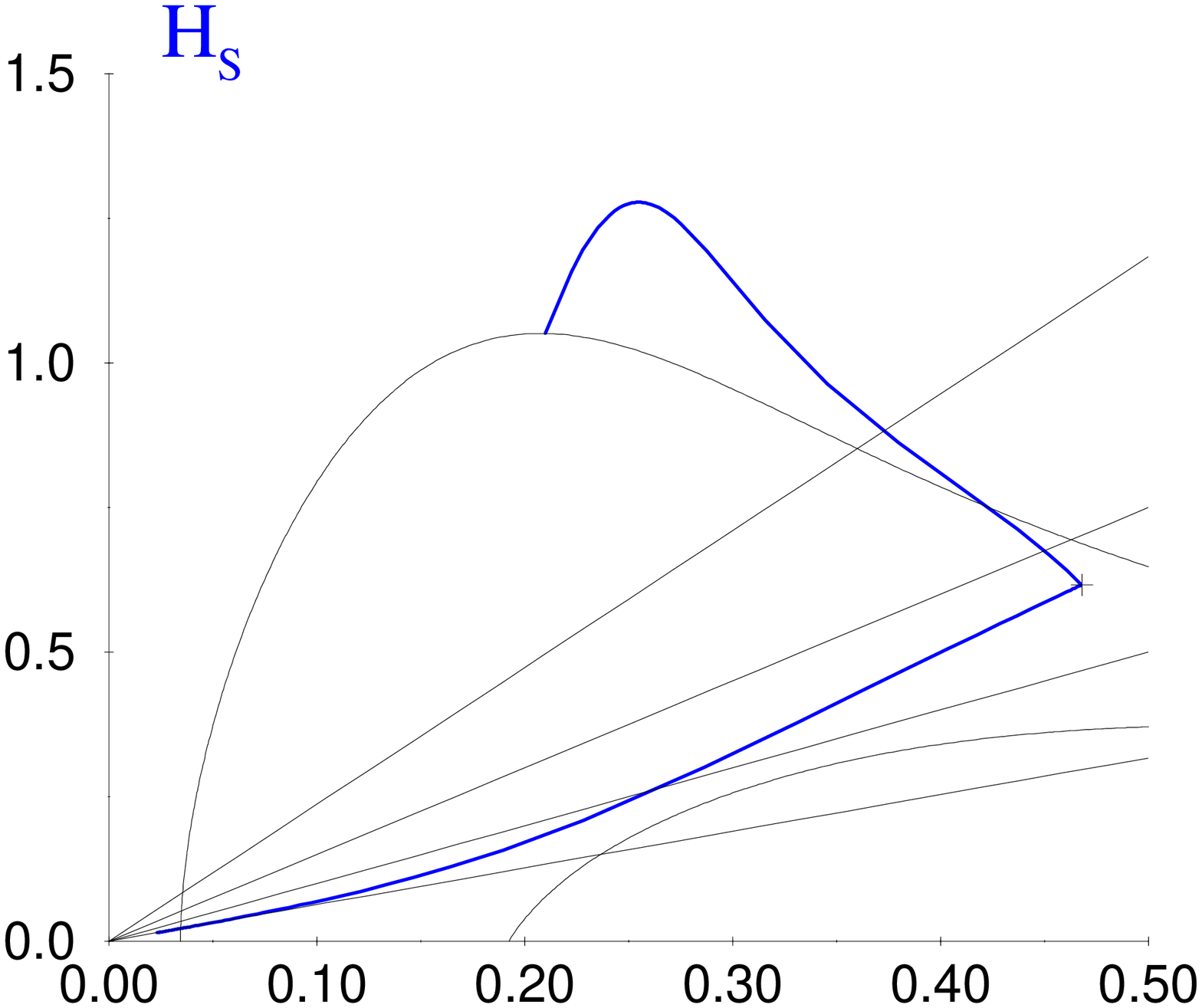} 
\caption 
{\small\it 
$\L_q=\L_q^{\phi}$, $\L_{m}=-0.02 \phi^2 e^{\phi}$. Quantities 
plotted and initial conditions are as in \figref{f:gmv}{a}. Also shown 
are the singularity curves, defined in the text. The right curve is 
for $\phi\rightarrow -\infty$ and the left curve is for the value of $\phi$
at which the solution hits the curve. } 
\label{f:gp4pot} 
\efig 
 
Concentrating on the simplest of the NEC violating quantum corrections, 
(\ref{gp4}), we are surprised to find obstacles to this program. Attempting 
to moderate the evolution through the use of a potential strong enough 
to affect the solution actually drives the solution into cusp  
singularities (singularities in $\ddot \phi$ and $\dot H_S$). 
The sources for a potential are:  
\begin{eqnarray} 
\half \L_m^{V}&=&V(\phi)\\ 
\rho_m^{V}&=&-V(\phi)\\ 
p_m^{V}&=&V(\phi)\\ 
\Delta_{\phi} \L_m^{V}&=&V'(\phi) 
\end{eqnarray} 
We illustrate a particular case in \figref{f:gp4pot}{},  
with $V(\phi)=-0.01 \phi^2 
e^{\phi}$. We have added graphs of the locations of the singularity 
curves. The curve on the right is the location of the singularities 
at $\phi \rightarrow -\infty$. The curve on the left is their location 
when the solution collides with them. Adding radiation can produce 
similar singularities or it is quickly redshifted away by the 
growing $H_S$. 
 
The source of these singularities is easy to understand mathematically. 
The sources introduce additional terms containing the highest 
derivatives $\dot H_S$ and $\ddot \phi$ to the terms coming from 
the lowest order terms in the action. When we solve the equations 
of motion for $\dot H_S$ and $\ddot \phi$ in terms of the lower  
derivatives we find a denominator which has, in general, zeros in the $(x,y)$ 
plane, where $x=\dot \phi/3$ and $y=H_S$. In this particular case this leads  
to the equation  for the singularity curves: 
\begin{equation} 
{1 \over 3} - 9\,{x^2} - 36\,{e^{\phi(t)}}\,{x^2} + 2\,x\,y + 27\,{x^3}\,y +  
108\,{e^{\phi(t)}}\,{x^3}\,y + 2\,{y^2} + \,{y^4}=0 
\end{equation} 
The terms containing the $e^\phi$ are coming from the quantum correction, 
so that the location of the singularity curves is now a function of $\phi$. 
 
Perversely, the singularity curves follow the solution towards small 
$\dot \phi$ and large $H_S$. So any attempt to tinker with the solution 
causes it to collide with them. Modeling the production of radiation 
produces a similar effect. While we do not have a direct physical 
interpretation of these singularities we do regard them as an indication 
of the general instability of NEC violating solutions. 
 
\bfig 
\bigeps{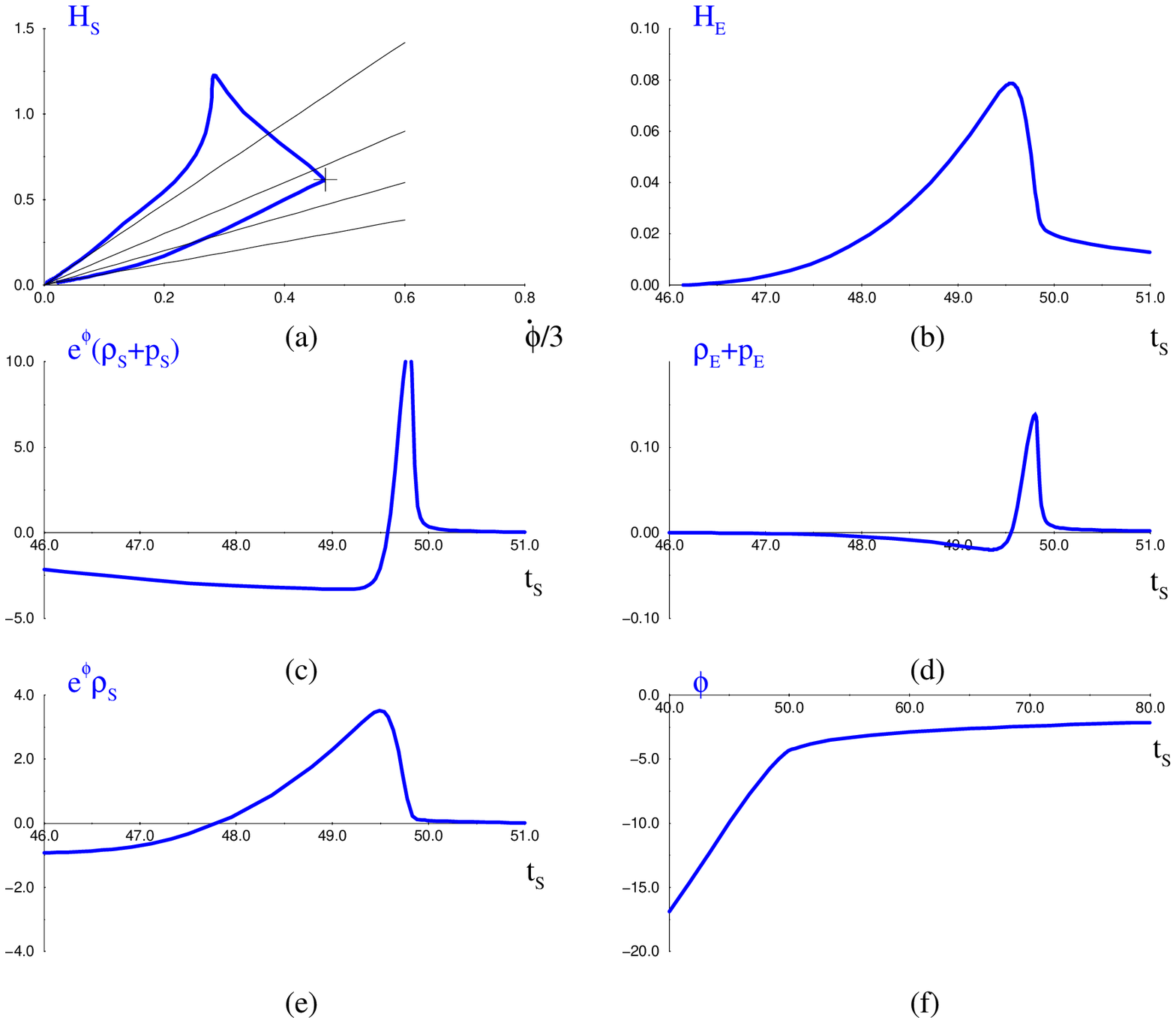}
\caption 
{\small\it 
$\L_q=\L_q^{f(\phi)}$, $f(\phi)=\half e^6 
(1-\tanh(8 \phi+36))$.  
See \figref{f:gmv}{} 
caption for details and initial conditions. All quantities are nonsingular,
but peaks are out of scale to emphasize details.} 
\label{f:dgp4} 
\efig 
 
A direct approach to completing the exit transition is to assume that there
exists some mechanism that  shuts off the correction terms, and hence, NEC
violation. A concrete way of modeling such an idea is simply replacing the 
quantum correction in the action with $f(\phi) \L_q^{\phi}$ where $f(\phi)$ 
is a positive
constant for $\phi<\phi_0$ for some constant $\phi_0$ 
and then smoothly goes to 
 zero, so $f(\phi)$ has the form of a smoothed step function. This
successfully eliminates the loop corrections at late  
times so the dilaton may be
easily captured by a potential or slowed by radiation production as shown in
\figref{f:dgp4}{}.  In this suite of figures we have dropped the branch sign
graph, since all of the following solutions are similar in that region of the
evolution and have now chosen a range of time to graph the sources emphasizing
the epoch when NEC violation ceases and the evolution 
becomes decelerated. We
have also tinkered with the normalization of this term so the figure may be
directly compared with the previous figures. 
While we have exactly the desired behavior, a function like $f(\phi)$
will not appear by summing a few terms in the loop expansion.
 
\begin{eqnarray} 
\half \L_q^{f(\phi)}&=& -f(\phi)(\nabla \phi)^4 \\ 
\rho_q^{f(\phi)}&=&-3 f(\phi) \dot \phi^4 \\ 
p_q^{f(\phi)}&=&-f(\phi) \dot \phi^4 \\ 
\Delta_{\phi} \L_q^{f(\phi)}&=& 
  12 f(\phi) H_S \dot \phi^3 +3 f'(\phi) \dot \phi^4 +  
   12 f(\phi) \dot \phi^2 \ddot \phi 
\end{eqnarray} 
 
\bfig 
\bigeps{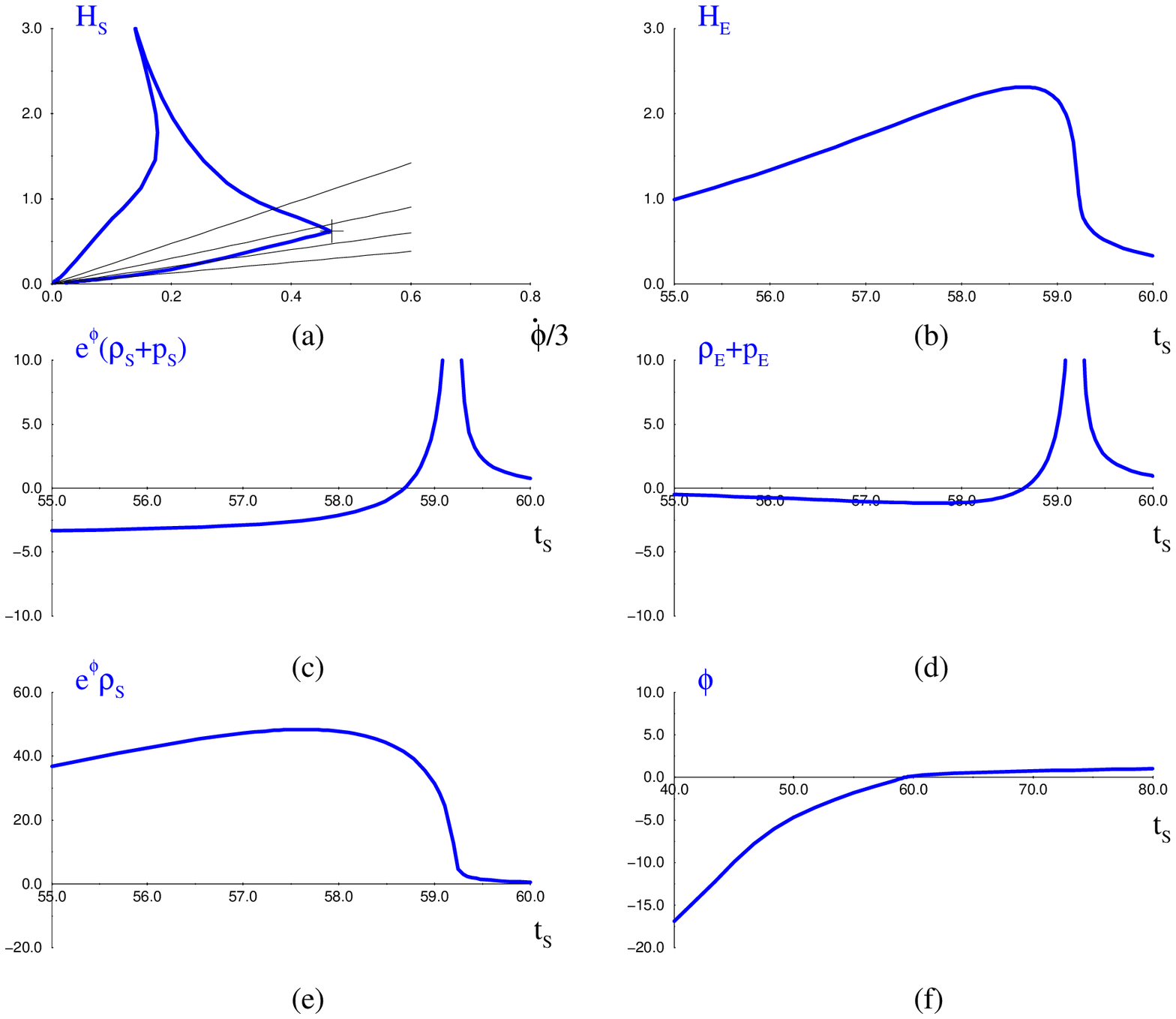}
\caption 
{\small\it 
$\L_q=1000 \L_q^{\phi}+1000 \L_q^{2 \phi}$. The peaks of  
$e^\phi(\rho_S+p_S) \simeq 55$ 
and $\rho_E+p_E \simeq 55$ have been cut off to emphasize the sign change,
but all quantities are non-singular.
See \figref{f:gmv}{} caption for details and initial conditions.} 
\label{f:gp42} 
\efig 
 
However, we have found that a complete suppression of the quantum 
corrections is not necessary. Suspecting that the instabilities 
are due to continued NEC violation we propose what is, perhaps, 
a simpler model. Since the NEC violation was  
induced in our models by one loop quantum corrections, 
higher loop terms can suppress 
the NEC violation once it is no longer needed if they
have the correct sign. For example, a two loop 
contribution of the form, 
\begin{eqnarray} 
\half \L_q^{2 \phi}&=&e^{\phi}(\nabla \phi)^4 \\ 
\rho_q^{2 \phi}&=&3 e^{\phi} \dot \phi^4 \\ 
p_q^{2 \phi}&=&e^{\phi} \dot \phi^4 \\ 
\Delta_{\phi} \L_q^{2 \phi}&=& 
  -e^{\phi} (12 H_S \dot \phi^3 + 3 \dot \phi^4+ 12\dot \phi^2 \ddot \phi) 
\end{eqnarray} 
can overwhelm the one loop NEC violation when $\phi$ becomes large
enough.
Since now the scaling of different terms with respect 
to a shift in the dilaton
(which determines the value of the string coupling at 
which various terms become 
important) is more complicated we will introduce these corrections with 
explicit large coefficients accounting for the expected large number of 
degrees of freedom contributing to the loop corrections. We expect, 
in string theory, large and approximately the 
same order of magnitude coefficients for 
certain one and two loop corrections, 
which, perhaps, may even be justified with some large N techniques. 
Taking $\L_q=C_1 \L_q^{\phi}+C_2 \L_q^{2 \phi}$ we observe that since 
the sources occur  in the equations of motion with coefficients 
$C_1 e^{\phi}$ and 
$C_2 e^{2 \phi}$ respectively, and these terms will be important 
to the evolution when these coefficients are of order unity, then
having $C_1 \approx C_2 \gg 1$ leads to a situation where 
the one-loop terms become important at a smaller value of $\phi$  
(and therefore an earlier time) than the two-loop. Thus we can 
still have an era of NEC violation which is ended by the onset of 
the two-loop terms in a rather natural way.
We have numerically solved the equations for a range of 
coefficients and observe a generic behaviour which 
we illustrate with the specific 
example in \figref{f:gp42}{}.
 
We show the results of a sample evolution in \figref{f:gp42}{}. For the graphs 
of $H_E$ and $\rho_E+p_E$ we choose a time range to 
emphasize the epoch after the bounce, 
where NEC violation ceases and the evolution becomes decelerated FRW. 
In \figref{f:gp42pot}{} we show that with  
this form of corrections the behavior is 
mild enough that it is easy to capture the dilaton 
into a potential minimum. 
In these figures we emphasize the late phase where the dilaton is rolling 
around the potential minimum, but the early evolution is indistinguishable 
from \figref{f:gp42}{}. 
 
\bfig 
\bigeps{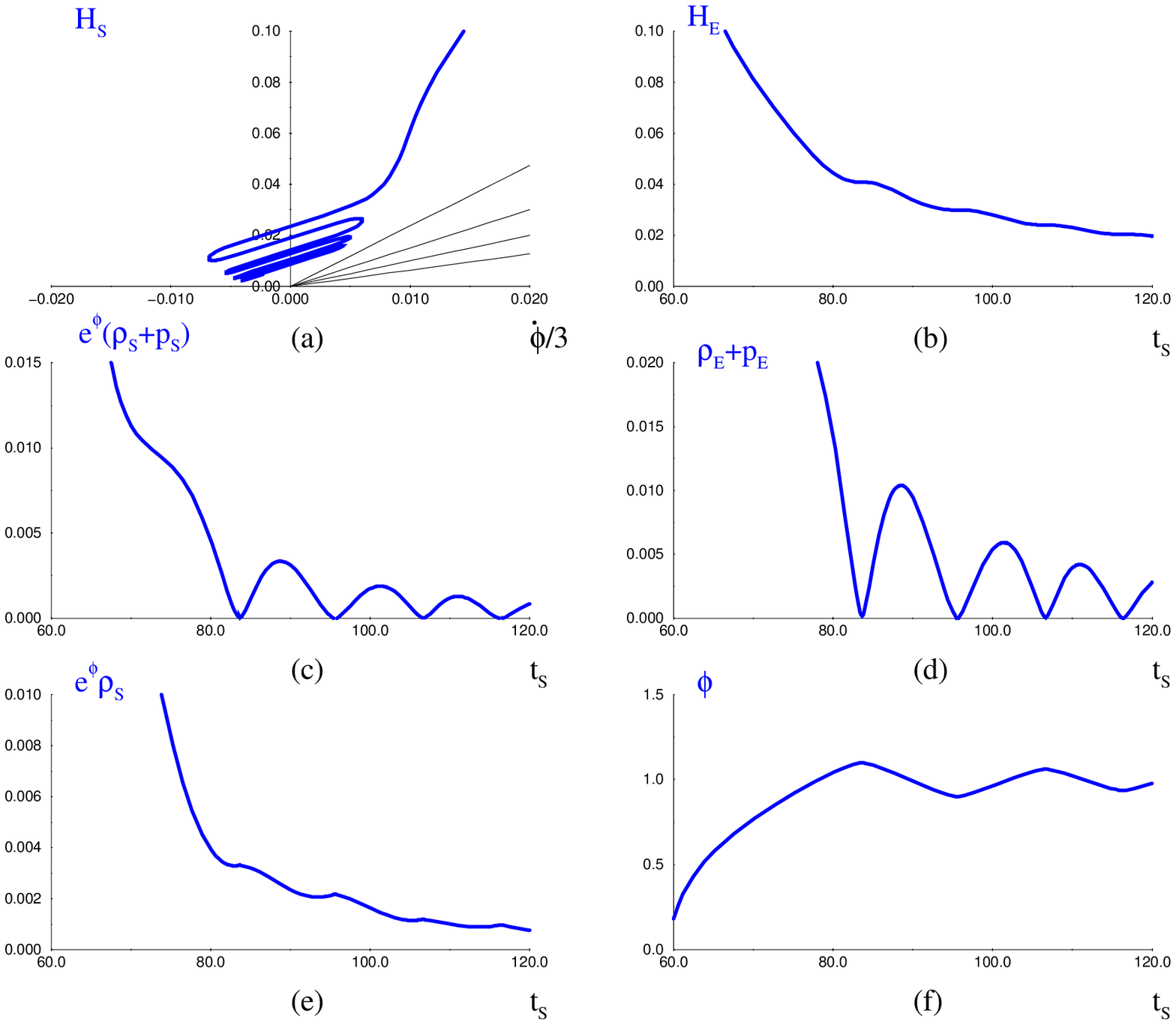}
\caption 
{\small\it 
$\half \L_q=1000 \L_q^{\phi}+1000 \L_q^{2 \phi},  
\half \L_m=-0.1(\phi-1)^2 e^{\phi-1}$. See \figref{f:gmv}{} 
caption for details and initial conditions. } 
\label{f:gp42pot} 
\efig 
 
We would also like to show that these solutions are stable enough
that the growing dilaton can be halted by introducing radiation,
and that they can pass into a radiation dominated phase and be
smoothly joined to standard cosmologies.
We search for the simplest consistent, but not necessarily realistic,
method of producing radiation. Since the radiation conservation equation 
(\ref{nconseq}) can be 
derived from the other three equations of motion, simply 
placing an arbitrary source into (\ref{nconseq}) is not satisfactory, 
and produces an inconsistent system.
Practically, since we are using the equations containing the highest 
derivatives to integrate the system, this means that when the radiation 
source turns on, we will begin violating the constraint equation 
(\ref{n00eq}). 

Instead we use the same ansatz used to  
model radiation production from the oscillation of the inflaton in 
slow-roll inflation models. We will produce the radiation from the 
dilaton kinetic energy. To do this we introduce a coupling of 
the radiation to the dilaton
by introducing a $\Delta_{\phi} \L_{rad}$ into 
(\ref{nconseq}) and the same term into (\ref{phieq}). The natural form to 
use is $\Delta_{\phi} \L_{rad} \propto \dot \phi$, since this will ensure our 
radiation source is non-negative. We repeat for emphasis that 
we do not claim this is the actual way radiation is produced, especially
since it violates the generic expectation that the dilaton will couple
to the trace of the energy momentum tensor which vanishes for radiation. 
The true source of radiation will likely consist of the particles 
produced abundantly in the dilaton-driven phase. But in the spirit of
this paper, it will serve to model the physics.

The results are shown in \figref{f:gp42rad}{}. 
The produced radiation slows the 
dilaton to a halt, in turn suppressing the corrections and creating 
a radiation dominated evolution as is shown in \figref{f:gp42rad}{d}.  
In \figref{f:gp42rad}{d} we have plotted both the total $\rho_E+p_E$  
and the contribution from the radiation alone $\rho_R+p_R$.  
The relation between radiation density 
in the string frame and Einstein frame is given by  
$\rho_E=e^{2 \phi}\rho_S$. 
 
\bfig 
\bigeps{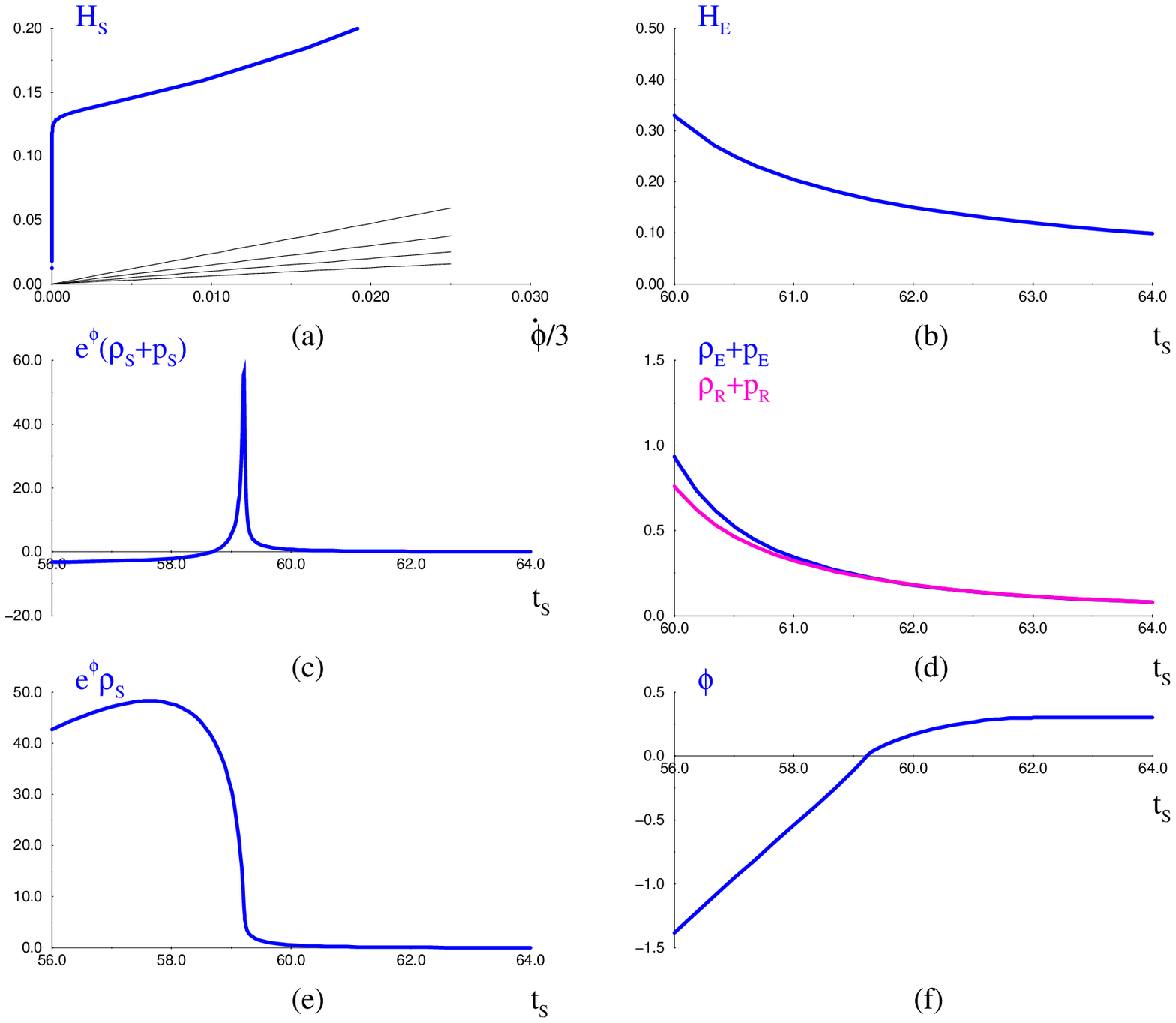}
\caption 
{\small\it 
$\L_q=1000 \L_q^{\phi}+1000 \L_q^{2 \phi},  
\Delta_{\phi} \L_{rad}=-10 \dot \phi$. See \figref{f:gmv}{} 
caption for details and initial conditions. We have superimposed 
$\rho_R+p_R$, the source quantities for the radiation onto 
\figref{f:gp42rad}{d} to show the onset of radiation domination.} 
\label{f:gp42rad} 
\efig 

These provide the first concrete examples of a completed graceful exit 
based on a classical evolution from an effective action. 
We have presented solutions interpolating between 
the inflationary $(+)$ branch to decelerated $(-)$ branch evolution 
in which the dilaton can be captured by a potential or stopped by 
radiation production. Using our analysis of the qualities and energy 
conditions required of sources to produce this transition we have  
arrived at an elusive destination.  
 
\section{Summary and Conclusions}

Graceful exit transition in string cosmology 
is not forbidden in principle as a variety of concrete 
examples show explicitly. 
We have verified the general arguments of \cite{BM}, 
showing that NEC violation 
is not just a necessary condition, but in some generic sense, also a 
sufficient condition. We have encountered yet another 
``exit problem" from a correction dominated evolution to a 
standard decelerated FRW evolution, which had to be overcome. We suggested 
that effective terms coming from higher loop corrections may do the job, 
and presented, for the first time, 
an effective Lagrangian whose equations of motion
possess non-singular solutions interpolating between a $(+)$ branch vacuum
and ordinary radiation dominated FRW evolution with a fixed dilaton.

The remaining questions concern whether specific string models produce 
coefficients of appropriate sign and size. We need to know if the
one-loop terms do indeed violate NEC and 
whether a physical shut-off mechanism
does operate in string theory. Note that the form of 
induced terms at one-loop is guaranteed, from general arguments,
to be the one we used.

Each of our non-singular string cosmology solutions provides
a detailed description of the high-curvature phase in between 
the dilaton-driven inflationary phase and the FRW decelerated 
phase (the ``string phase"). In all solutions we find that a 
phase rich in structure appears which is much more complicated
than the one postulated in \cite{bggv} of constant $\dot\phi$ and
$H_S$. We do not suggest taking the details of the 
evolution in the string phase too
seriously, because the terms that determine those details have arbitrary
coefficients. However, our examples could be taken 
as an illustration of what the
real string phase may eventually look like.

\acknowledgements

We thank G. Veneziano for discussions about the coefficients of quantum
correction terms and comments on the manuscript, for which we are also thankful
to J. Maharana. This work is supported in part by the  Israel Science
Foundation administered by the Israel Academy of Sciences and Humanities.

\appendix
\section*{Effective terms, equations of motion and their solutions}
\label{appA}

Before embarking on determining whether a definite string 
model has correction terms of the form required to induce the graceful exit,
we have made this survey of likely forms of source terms. To do
this required the variation of many forms of Lagrangian terms with 
many combinations of coefficients. Our interest was not in any one 
specific action. To make this feasible we have developed
software to handle many of the aspects of the journey from action, through 
equations of motion and numerical integration and finally to graphical
representation of the resulting dynamics in a useful form.

The core of the process is the automated derivation of the 
equations of motion by
varying the action, in a form suitable for numerical integration. 
Making the    
process difficult was the requirement that we handle essentially 
any form of    
correction term in the action. But making
the prospect easier was our specializing on homogeneous and isotropic
solutions and the use of a tireless symbol manipulator, 
{\it Mathematica} \cite{mathematica}. 
This enables us to use an almost embarrassingly blunt approach.
First we construct a matrix to represent the metric, in our case,
$diag(g_{00}(t),g_{11}(t,r),g_{22}(t,r),g_{33}(t,r,\phi))$. 
Then we construct 
the tensor quantities we need in the most direct possible way. We use
the metric to compute the Christoffel symbols and proceed to the
Riemann tensor and contractions thereof, all of which are stored in
lists. From here we compute any required geometrical scalars for the
action, which all emerge in a raw form, 
completely in terms of $g_{00}...g_{33}$ and
their various partial derivatives and any other fields.

From here we construct the $\rho$ and $p$ and other quantities by
directly varying with respect to the metric fields and other
quantities. Our techniques for doing this were helped greatly
by study of \cite{yh}. Then we put in the cosmic time gauge 
choice $g_{00}=-1$ and the components of the usual FRW metric 
$a(t)^2 d\Omega$ for $g_{11},g_{22},g_{33}$. Finally we replace the derivatives
of $a(t)$ with their corresponding expressions in terms of the Hubble
parameter $H=\dot a/a$.

The very crudeness of this process enhances our belief in its correctness.
Other consistency checks are possible, e.g. we make the redundant check
that $T^1_1=T^2_2=T^3_3$. We also verify the conservation equation 
(\ref{nconseq}) for these sources. They replicate many known
results, e.g. the vanishing of sources from the one-loop Gauss-Bonnet
combination, and we can reproduce the numerical integration of other
examples in the literature. On a case by case basis 
we check the accuracy of numerical integrations by 
verifying the constraint equation (\ref{n00eq}).

Finally, we present a table of generalized sources, sufficient to
construct all of the equations of motion used in this work, a generalized
form of the dilaton kinetic term (\ref{agendp}), the Ricci scalar with
arbitrary dilaton dependence (\ref{agenr}) and various $R^2$ combinations
(\ref{agenr2}, \ref{agenr22} and \ref{agenr24}). $k$ is the sign of the
spatial curvature.

\begin{eqnarray}
{\cal{L}}_{q}^{g((\nabla \phi)^ 2)}&=&f(\phi)g((\nabla \phi)^ 2)
\label{agendp}\\
\nonumber \rho_{q}^{g((\nabla \phi)^ 2)}&=&
{\frac{-\left( f(\phi)g((\nabla \phi)^ 2) \right) }{2}} - 
f(\phi)g'((\nabla \phi)^ 2){{\dot \phi}^2}\\
\nonumber p_{q}^{g((\nabla \phi)^ 2)}&=&
{\frac{f(\phi)g((\nabla \phi)^ 2)}{2}}\\
\nonumber \Delta_\phi {\cal{L}}_{q}^{g((\nabla \phi)^ 2)}&=&
{\frac{g((\nabla \phi)^ 2)f'(\phi)}{2}} + 
3f(\phi)Hg'((\nabla \phi)^ 2)\dot \phi + 
f'(\phi)g'((\nabla \phi)^ 2){{\dot \phi}^2} + 
f(\phi)g'((\nabla \phi)^ 2)\ddot \phi - 
2f(\phi){{\dot \phi}^2}g''((\nabla \phi)^ 2)\ddot \phi\\
\nonumber ({{(\nabla \phi)}^ 2}) &=& -{{\dot \phi}^2}
\end{eqnarray}

\begin{eqnarray}
{\cal{L}}_{q}^{R}&=&R f(\phi)
\label{agenr}\\
\nonumber \rho_{q}^{R}&=&{\frac{-3kf(\phi)}{{{a(t)}^2}}} - 
3f(\phi){H^2} - 3Hf'(\phi)\dot \phi\\
\nonumber p_{q}^{R}&=&{\frac{kf(\phi)}{{{a(t)}^2}}} + 
3f(\phi){H^2} + 2f(\phi)\dot H + 2Hf'(\phi)\dot \phi + 
{{\dot \phi}^2}f''(\phi) + f'(\phi)\ddot \phi\\
\nonumber \Delta_\phi {\cal{L}}_{q}^{R}&=&
{\frac{3kf'(\phi)}{{{a(t)}^2}}} + 6{H^2}f'(\phi) + 
3f'(\phi)\dot H
\end{eqnarray}

\begin{eqnarray}
{\cal{L}}_{q}^{{R^2}}&=&R^2 f(\phi)
\label{agenr2}\\
\nonumber \rho_{q}^{{R^2}}&=&{\frac{-18{k^2}f(\phi)}{{{a(t)}^4}}} + 
{\frac{36kf(\phi){H^2}}{{{a(t)}^2}}} - 108f(\phi){H^2}\dot H + 
18f(\phi){{\dot H}^2} - 
{\frac{36kHf'(\phi)\dot \phi}{{{a(t)}^2}}} - 
72{H^3}f'(\phi)\dot \phi - \\
\nonumber & & 36Hf'(\phi)\dot H\dot \phi - 
36f(\phi)H\ddot H\\
\nonumber p_{q}^{{R^2}}&=&{\frac{-6{k^2}f(\phi)}{{{a(t)}^4}}} - 
{\frac{12kf(\phi){H^2}}{{{a(t)}^2}}} - 
{\frac{24kf(\phi)\dot H}{{{a(t)}^2}}} + 
108f(\phi){H^2}\dot H + 54f(\phi){{\dot H}^2} - 
{\frac{24kHf'(\phi)\dot \phi}{{{a(t)}^2}}} + 
48{H^3}f'(\phi)\dot \phi + \\
\nonumber & & 120Hf'(\phi)\dot H\dot \phi + 
{\frac{12k{{\dot \phi}^2}f''(\phi)}{{{a(t)}^2}}} + 
24{H^2}{{\dot \phi}^2}f''(\phi) + 
12\dot H{{\dot \phi}^2}f''(\phi) + 72f(\phi)H\ddot H + 
24f'(\phi)\dot \phi\ddot H + 
{\frac{12kf'(\phi)\ddot \phi}{{{a(t)}^2}}} + \\
\nonumber & &
24{H^2}f'(\phi)\ddot \phi + 12f'(\phi)\dot H\ddot \phi + 
12f(\phi)H^ {(3)}\\
\nonumber \Delta_\phi {\cal{L}}_{q}^{{R^2}}&=&
{\frac{18{k^2}f'(\phi)}{{{a(t)}^4}}} + 
{\frac{72k{H^2}f'(\phi)}{{{a(t)}^2}}} + 
72{H^4}f'(\phi) + {\frac{36kf'(\phi)\dot H}{{{a(t)}^2}}} + 
72{H^2}f'(\phi)\dot H + 18f'(\phi){{\dot H}^2}
\end{eqnarray}

\begin{eqnarray}
{\cal{L}}_{q}^{R^ {\mu \nu} R_{\mu \nu}}&=&
R^ {\mu \nu} R_{\mu \nu} f(\phi) \label{agenr22}\\
\nonumber \rho_{q}^{R^ {\mu \nu} R_{\mu \nu}}&=&
{\frac{-6{k^2}f(\phi)}{{{a(t)}^4}}} + 
{\frac{12kf(\phi){H^2}}{{{a(t)}^2}}} - 
36f(\phi){H^2}\dot H + 6f(\phi){{\dot H}^2} - 
{\frac{6kHf'(\phi)\dot \phi}{{{a(t)}^2}}} - 
18{H^3}f'(\phi)\dot \phi - \\
\nonumber & & 12Hf'(\phi)\dot H\dot \phi - 
12f(\phi)H\ddot H\\
\nonumber p_{q}^{R^ {\mu \nu} R_{\mu \nu}}&=&
{\frac{-2{k^2}f(\phi)}{{{a(t)}^4}}} - 
{\frac{4kf(\phi){H^2}}{{{a(t)}^2}}} - 
{\frac{8kf(\phi)\dot H}{{{a(t)}^2}}} + 36f(\phi){H^2}\dot H + 
18f(\phi){{\dot H}^2} - 
{\frac{8kHf'(\phi)\dot \phi}{{{a(t)}^2}}} + 
12{H^3}f'(\phi)\dot \phi + \\
\nonumber & & 36Hf'(\phi)\dot H\dot \phi + 
{\frac{2k{{\dot \phi}^2}f''(\phi)}{{{a(t)}^2}}} + 
6{H^2}{{\dot \phi}^2}f''(\phi) + 
4\dot H{{\dot \phi}^2}f''(\phi) + 24f(\phi)H\ddot H + 
8f'(\phi)\dot \phi\ddot H + \\
\nonumber & &
{\frac{2kf'(\phi)\ddot \phi}{{{a(t)}^2}}} + 
6{H^2}f'(\phi)\ddot \phi + 4f'(\phi)\dot H\ddot \phi + 
4f(\phi)H^ {(3)}\\
\nonumber \Delta_\phi {\cal{L}}_{q}^{R^ {\mu \nu} R_{\mu \nu}}&=&
{\frac{6{k^2}f'(\phi)}{{{a(t)}^4}}} + 
{\frac{18k{H^2}f'(\phi)}{{{a(t)}^2}}} + 18{H^4}f'(\phi) + 
{\frac{6kf'(\phi)\dot H}{{{a(t)}^2}}} + 18{H^2}f'(\phi)\dot H + 
6f'(\phi){{\dot H}^2}
\end{eqnarray}

\newpage
\begin{eqnarray}
{\cal{L}}_{q}^{R^ {\mu \nu \sigma \lambda} 
R_{\mu \nu \sigma \lambda}}&=&
R^ {\mu \nu \sigma \lambda} R_{\mu \nu \sigma \lambda} f(\phi)
\label{agenr24}\\
\nonumber \rho_{q}^{R^ {\mu \nu \sigma \lambda} R_{\mu \nu \sigma \lambda}}
&=&{\frac{-6{k^2}f(\phi)}{{{a(t)}^4}}} + 
{\frac{12kf(\phi){H^2}}{{{a(t)}^2}}} - 
36f(\phi){H^2}\dot H + 6f(\phi){{\dot H}^2} - 
12{H^3}f'(\phi)\dot \phi - 12Hf'(\phi)\dot H\dot \phi - 
12f(\phi)H\ddot H\\
\nonumber p_{q}^{R^ {\mu \nu \sigma \lambda} R_{\mu \nu \sigma \lambda}}&=&
{\frac{-2{k^2}f(\phi)}{{{a(t)}^4}}} - 
{\frac{4kf(\phi){H^2}}{{{a(t)}^2}}} - 
{\frac{8kf(\phi)\dot H}{{{a(t)}^2}}} + 36f(\phi){H^2}\dot H + 
18f(\phi){{\dot H}^2} - 
{\frac{8kHf'(\phi)\dot \phi}{{{a(t)}^2}}} + 
8{H^3}f'(\phi)\dot \phi + \\
\nonumber & & 32Hf'(\phi)\dot H\dot \phi + 
4{H^2}{{\dot \phi}^2}f''(\phi) + 4\dot H{{\dot \phi}^2}f''(\phi) 
+ 24f(\phi)H\ddot H + 8f'(\phi)\dot \phi\ddot H 
+ 4{H^2}f'(\phi)\ddot \phi + \\
\nonumber & & 4f'(\phi)\dot H\ddot \phi + 
4f(\phi)H^ {(3)}\\
\nonumber \Delta_\phi {\cal{L}}_{q}^{R^ {\mu \nu \sigma \lambda} 
R_{\mu \nu \sigma \lambda}}&=&{\frac{6{k^2}f'(\phi)}{{{a(t)}^4}}} + 
{\frac{12k{H^2}f'(\phi)}{{{a(t)}^2}}} + 12{H^4}f'(\phi) + 
12{H^2}f'(\phi)\dot H + 6f'(\phi){{\dot H}^2}
\end{eqnarray}

\end{document}